\documentclass[prl,twocolumn]{revtex4}
\usepackage{graphicx}
\usepackage{amssymb,amsmath,bm}
\usepackage{enumerate}

\newcommand{\be}{\begin{eqnarray}}
\newcommand{\ee}{\end{eqnarray}}

\begin{document}

\title{Distributed chaos and Rayleigh-B\'{e}nard turbulence at very high Ra}

\author{A. Bershadskii}

\affiliation{
ICAR, P.O. Box 31155, Jerusalem 91000, Israel
}

\begin{abstract}

It is shown, by the means of distributed chaos approach and using the experimental data, that at very large Rayleigh number $Ra > 10^{14}$ and Prandtl number $Pr \sim 1$ the Rayleigh-B\'{e}nard turbulence can undergo a transition related to spontaneous breaking of the fundamental Lagrangian relabeling symmetry. Due to the Noether's theorem helicity plays central role in this process. After the transition the temperature spectrum has a stretched exponential form $E (k) \propto \exp(-k/k_{\beta})^{\beta}$ with $\beta =2/5$ both at the cell midplain and at the near-wall (low boundary) regions. There is a similarity between this phenomenon and the effects of polymer additives.

\end{abstract}

\maketitle

\section{Inroduction}

The Rayleigh-B\'{e}nard convection (RBC) in the Boussinesq approximation can be described by the standard equations used in the theory of buoyancy driven motions with imposed linear temperature gradient \cite{bur}
$$ 
\frac{ \partial{\mathbf u}}{\partial t} +{\mathbf u} \cdot \nabla {\mathbf u}  = -\frac{1}{\rho_{0}} \nabla P - N \theta  {\bf e_g} + \nu \nabla^2 {\mathbf u}+{\bf f}  \eqno{(1)}, 
$$ 
$$     
\frac {\partial \theta}{\partial t} +{\mathbf u} \cdot \nabla \theta  = - N~ {\bf u} \cdot {\bf e_z} + D \nabla^2 \theta \eqno{(2)},
$$  
$$
 \nabla \cdot {\bf u} = 0 \eqno{(3)};
$$
if temperature $T$ is replaced by $\theta = T - \Delta T \cdot z/L$ (pressure replacement should be also made in order to compensate an additional term arising in the Eq. (1)). Here and further we will consider situations when condition ${\bf e_g} = {\bf e_z}$ is satisfied with sufficient precision (see Ref. \cite{b}). At this replacement $L$ is the convection cell height, $T_{l}$ is temperature of the cell low boundary, $T_{l} + \Delta T$ is temperature of the upper boundary and $\theta = 0$ for the upper and low boundaries. The new variable $\theta$ is rescaled as a velocity. Corresponding Brunt-V\"ais\"al\"a frequency $N=\sqrt{g\delta \Delta T/H}$ (where $\delta$ is the thermal expansion coefficient and $g$ is the gravity acceleration). \\

The space translational symmetry (homogeneity) is related by the Noether's theorem to the momentum conservation \cite{ll2}.  The Birkhoff-Saffman integral 
$$
I_2 =\int  \langle {\bf u} \cdot  {\bf u'} \rangle d{\bf r} \eqno{(4)}
$$   
(where ${\bf u'} ={\bf u} ({\bf x} + {\bf r},t) $ and ${\bf u} = {\bf u} ({\bf x},t)$) is an invariant of the isotropic homogeneous Navier-Stokes equations and this invariant corresponds to the momentum conservation \cite{saf},\cite{dav1},\cite{dav2}.  Distributed chaos in isotropic homogeneous turbulence is dominated by the Birkhoff-Saffman integral that results in the spectrum
$$
E (k) \propto \exp(-k/k_{\beta})^{\beta}  \eqno{(5)}
$$
with $\beta =3/4$ \cite{b1}. An asymptotic scaling of the group velocity $\upsilon (\kappa )$ of the waves driving the distributed chaos
$$
\upsilon (\kappa )\propto I_2^{1/2}~\kappa^{3/2} \eqno{(6)}
$$
and relation 
$$
\beta =\frac{2\alpha}{1+2\alpha}   \eqno{(7)}
$$ 
was used in the Ref \cite{b1} in order to obtain this value of $\beta$ from the dimensional considerations.

 A buoyancy generalization of the Birkhoff-Saffman integral
$$
I_b =   \int  \langle {\bf u} \cdot  {\bf u'} - \theta~\theta' \rangle  d{\bf r}  \eqno{(8)}  
$$ 
allows to extend this considerations on the buoyancy driven turbulence \cite{b}. 

\section{Spontaneous breaking of the relabeling symmetry}
 
   For weak turbulence a spontaneous breaking of the space translational symmetry (homogeneity) was considered for the first time in Ref. \cite{nrz}. For strong turbulence (the Navier-Stokes dynamics) a theory of spontaneous breaking of the space translational symmetry by the viscosity and finite boundary conditions was suggested in recent Ref. \cite{b2}. In this theory
the distributed chaos is dominated by vorticity correlation integral
$$
\gamma = \int_{V} \langle {\boldsymbol \omega} \cdot  {\boldsymbol \omega'} \rangle_{V}  d{\bf r} \eqno{(9)}. 
$$  
where the vorticity ${\boldsymbol \omega} = \nabla \times {\bf u}$.
Substitution the parameter $\gamma$ into the Eq. (6) (instead of $I_2$) results in
$$
\upsilon (\kappa ) \propto |\gamma|^{1/2}~\kappa^{1/2} \eqno{(10)}
$$
and, then, in $\beta =1/2$. Direct generalization of this theory for the buoyancy driven turbulence with replacement of the parameter $\gamma$ by
$$
\gamma_b = \int_{V} \langle {\boldsymbol \omega} \cdot  {\boldsymbol \omega'} - Pr^{-1} \nabla \theta \cdot \nabla \theta' \rangle_{V}  d{\bf r} \eqno{(11)}. 
$$  
($Pr= \nu/D$ is the Prandtl number) results in the same value $\beta = 1/2$ in this case as well. Now both the viscosity $\nu \nabla^2 {\mathbf u}$ and diffusivity $D \nabla^2 \theta$ terms in Eqs. (1-2) together with the finite boundary conditions determine the spontaneous breaking of the space translational symmetry (homogeneity) \cite{b}. 

   Therefore, Prandtl number is a significant parameter for this phenomenon. If, for instance $Pr \gg 1$, then it follows from the Eq. (11) that the spontaneous breaking of the space translational symmetry (homogeneity) can determine the distributed chaos even at very large $Ra$ (it is rather possible that also the condition $1 \gg Pr$ has the same effect). An experimentally obtained temperature spectrum has been shown in Figure 1 in order to confirm this conclusion. The data were taken from Ref. \cite{as} for Rayleigh number $Ra = 3\cdot 10^{14}$ (cf next Section) and Prandtl number $Pr =300$. The stretched exponential spectral law (Eq. (5)) with $\beta = 1/2$ has been indicated by the dashed line. The Taylor hypothesis \cite{my} was used in order to transform the the frequency spectra into wavenumber spectra.\\

  However for $Pr \sim 1 $  a more complex phenomenon can take place. The viscosity and diffusivity terms together with the finite boundary conditions and with the buoyancy term $N \theta  {\bf e_g}$ (in Eq. (1)) can result in spontaneous breaking of the Lagrangian relabeling symmetry \cite{mas}, instead of the space translational symmetry. For the Euler equations this symmetry results (due to the Noether's theorem) in helicity 
$$
\mathcal{H} = \int  h ({\bf r},t) d{\bf r}  \eqno{(12)}
$$  
conservation (here $h={\bf u}\cdot{\boldsymbol \omega}$ ) \cite{y},\cite{pm},\cite{fs}. Namely,
$$
\frac{d\mathcal{H}}{dt}=\mathrm{Vi} + \mathrm{Bu}=\gamma_h \eqno{(13)}
$$  
where $\mathrm{Bu}$ and $\mathrm{Vi}$ are the buoyancy and viscous terms, correspondingly. The diffusivity term $D \nabla^2 \theta$ participates in the $\gamma_h$ through the buoyancy term $Bu$. In the vein of the Ref. \cite{b2} we will replace the parameter $\gamma$ in the Eq. (10) by the parameter $\gamma_h$ and will obtain from the dimensional considerations
$$
\upsilon (\kappa ) \propto |\gamma_h|^{1/3}~\kappa^{1/3} \eqno{(14)}
$$ 
and from Eq. (7) $\beta = 2/5$.  \\

\begin{figure}
\begin{center}
\includegraphics[width=8cm \vspace{-1cm}]{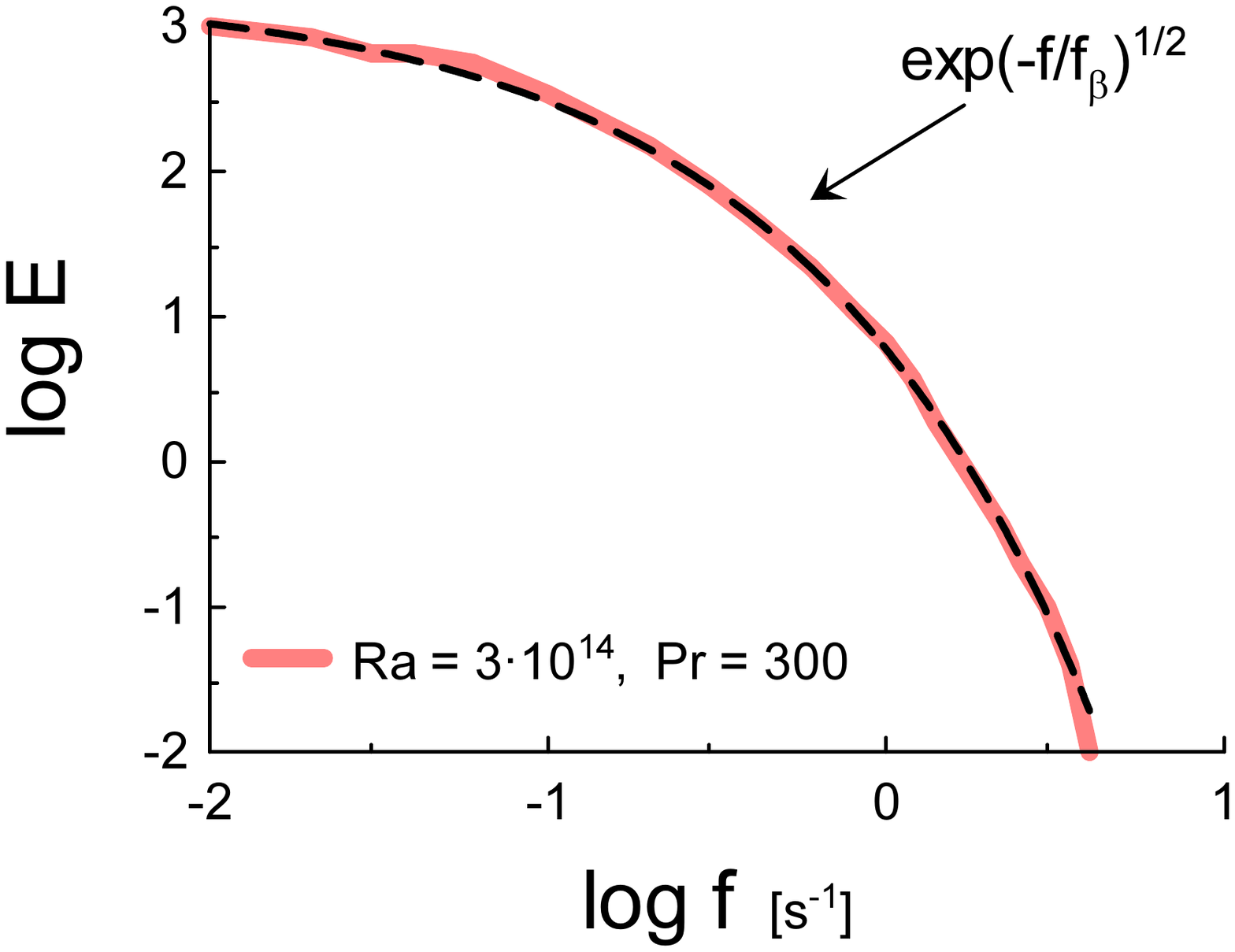}\vspace{-3.5cm}
\caption{\label{fig1} Temperature spectrum measured at $Ra = 3\cdot 10^{14}$ and Prandtl number $Pr =300$. The dashed line indicates the stretched exponential spectral law Eq. (5) with $\beta = 1/2$. }
\end{center}
\end{figure}
  It should be noted that the situation resembles the case of the distributed chaos in turbulence with polymer additives \cite{b3} (see also Discussion). 
   \begin{figure}
\begin{center}
\includegraphics[width=8cm \vspace{-1cm}]{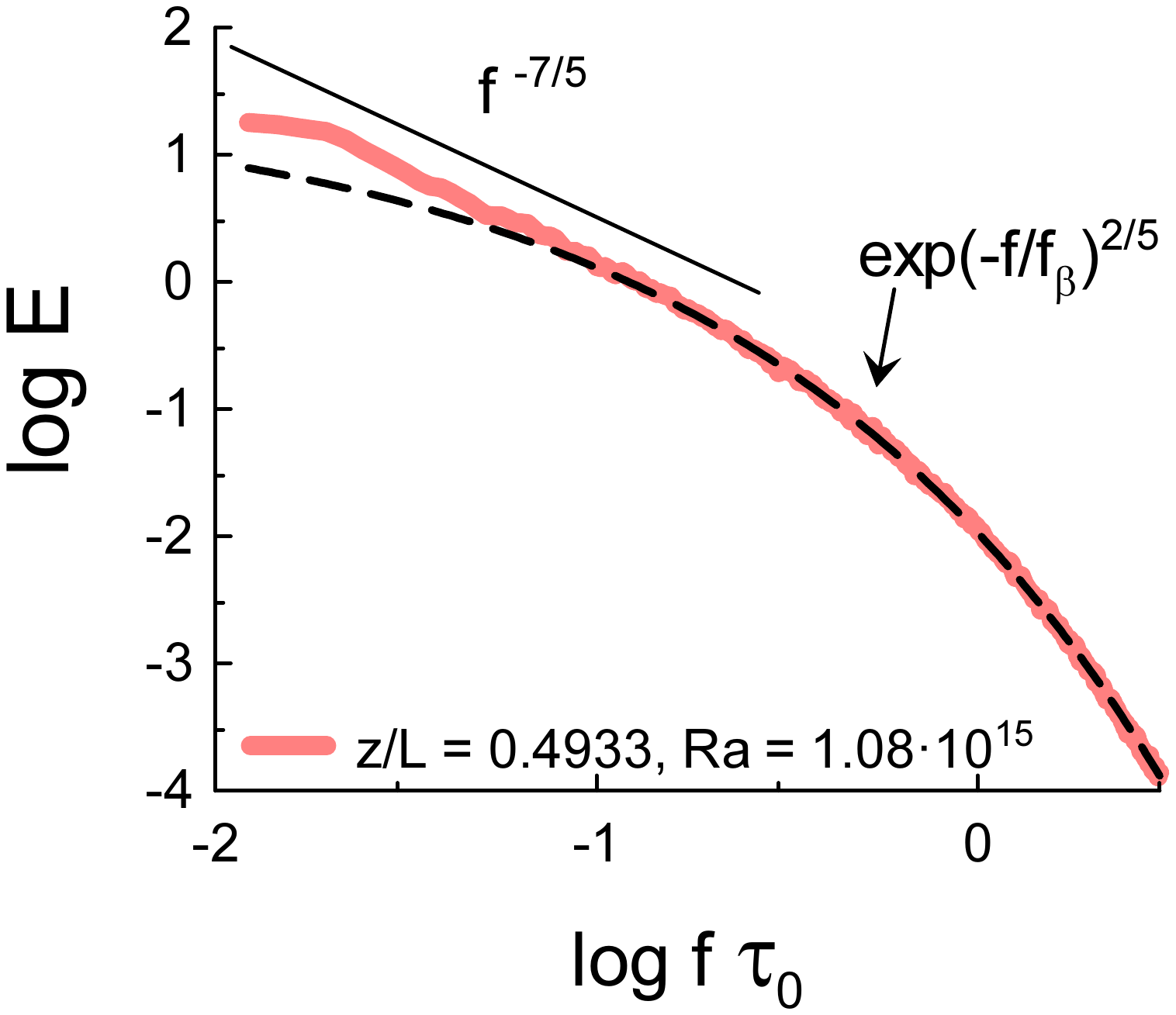}\vspace{-3.5cm}
\caption{\label{fig2} Temperature spectrum measured at $Ra = 1.08 \times 10^{15}$ and Prandtl number $Pr =0.8$ near the horizontal midplane. The dashed line indicates the spectral law Eq. (5) with $\beta = 2/5$. }
\end{center}
\end{figure}

\begin{figure}
\begin{center}
\includegraphics[width=8cm \vspace{-1.6cm}]{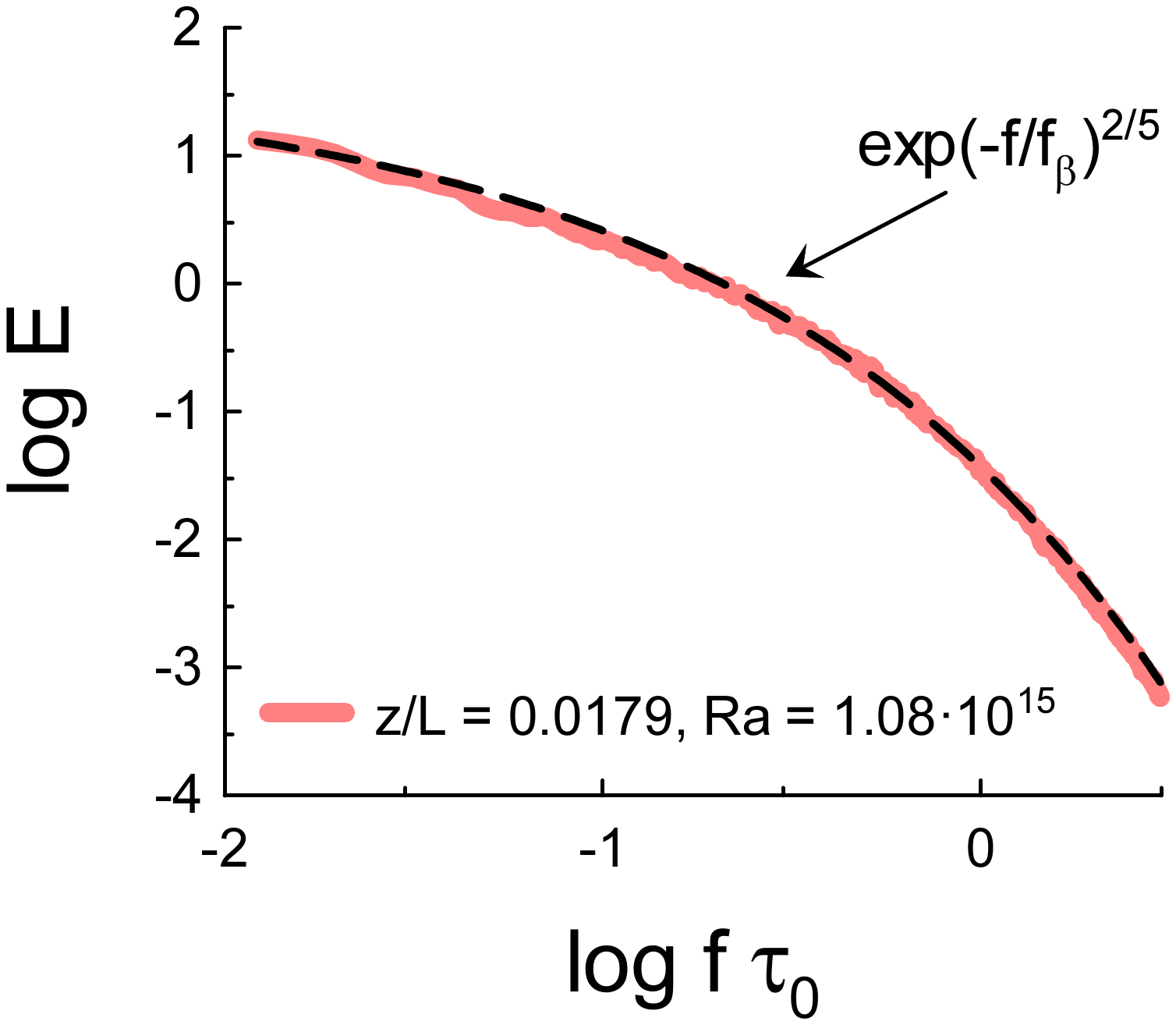}\vspace{-2.3cm}
\caption{\label{fig3} The same as in Fig.  2 but at the near-wall region. The dashed line indicates the spectral law Eq. (5) with $\beta = 2/5$. }. 
\end{center} 
\end{figure}

\section{Comparison with the experimental data at ${\boldsymbol Pr \sim 1}$}   

   It is believed that at Rayleigh number $Ra \sim 10^{14}$ the thin (but significant for the temperature dynamics) boundary layers near the low and upper boundaries become turbulent and the global transport properties of the Rayleigh-B\'{e}nard convection are significantly changing (despite that the bulk of the fluid in the cell was already highly turbulent for much smaller values of $Ra$). This change indicates a type of transition from one global regime (so-called 'classical') to another (so-called 'ultimate'). In order to understand dynamic nature of this transition it is useful to look at local fundamental quantities such as spectra, for instance. While the spectral properties were studied in detail (at least experimentally) for the classic regime, the experimental studies of the corresponding properties for the ultimate regime are in their embryonic stage. Therefore results of a recent experiment with $Ra = 1.08 \times 10^{15}$ ($Pr =0.8$) reported in Ref. \cite{he} are so interesting. It is also important that the temperature spectral data were obtained both at near the horizontal midplane ($z/L \simeq 0.4933$) and at the near-wall region ($z/L = 0.0179$). The radial location of the probes is $(R-r)/2R = 0.0178$. Global circulation (that indicates itself by a peak in the spectrum \cite{n}) was not observed at this experiment with the aspect ratio $2R/L = 1/2$ \cite{he1}. 
   
   It should be also noted that for these laboratory conditions a gradual transition from the classical regime to another one was observed in range $10^{13} < Ra < 5\cdot 10^{14}$ \cite{he2}. In the transitional range of $Ra$ (where a multistability was observed) applicability of the distributed chaos approach is rather questionable.\\
 
   Figure 2 shows the temperature spectrum measured at near the horizontal midplane in the log-log scales ($Ra = 1.08 \times 10^{15}$). The frequency was normalized by $\tau_0 = \sqrt{2R_0}$, where $R_0$ is the curvature radius of the temperature autocorrelation function at $\tau=0$ (details about relevance of this normalizaion can be found in the Ref. \cite{he}). The Taylor hypothesis transforming the frequency spectra into wavenumber spectra was shown to be valid at this experiment \cite{he}. 
   The dashed line is drawn in order to indicate the spectrum Eq. (5) with $\beta =2/5$ (i.e. spectrum corresponding to the distributed chaos with spontaneous breaking of the relabeling symmetry). The solid straight line indicates a possibility of the Bolgiano-Obukhov scaling  spectrum at small wavenumbers $k$ \cite{pz},\cite{bnps}. \\
    
    Figure 3 shows analogous spectrum obtained at the near-wall region. The dashed line is drawn in order to indicate the spectrum Eq. (5) with $\beta =2/5$. Just in the near-wall region the spectrum corresponding to the distributed chaos with spontaneous breaking of the relabeling symmetry covers the entire region of the scales. This fact supports the idea that the turbulization of the boundary layers is the main condition for the above mentioned global transition, related to the spontaneous breaking of the relabeling symmetry.

\section{Discussion}

  As for the spontaneous breaking of the translational symmetry the time dependence of the velocity correlation integral Eq. (2) (or the generalized momentum correlation integral Eq. (8)) should be considered, so for the spontaneous breaking of the relabeling symmetry the time dependence of the helicity correlation integral
$$
I = \int_V \langle h~h'\rangle_V d{\bf r} \eqno{(15),}
$$
should be considered in the same vein. However, it can be readily shown that in this case $\alpha =\beta =0$. Therefore, we have considered the helicity time dependence Eq. (13) instead (both the helicity $\mathcal{H}$ and the helicity correlation integral $I$ Eq. (15) are invariants of the Euler equation \cite{lt}).  \\

\begin{figure}
\begin{center}
\includegraphics[width=8cm \vspace{-1.15cm}]{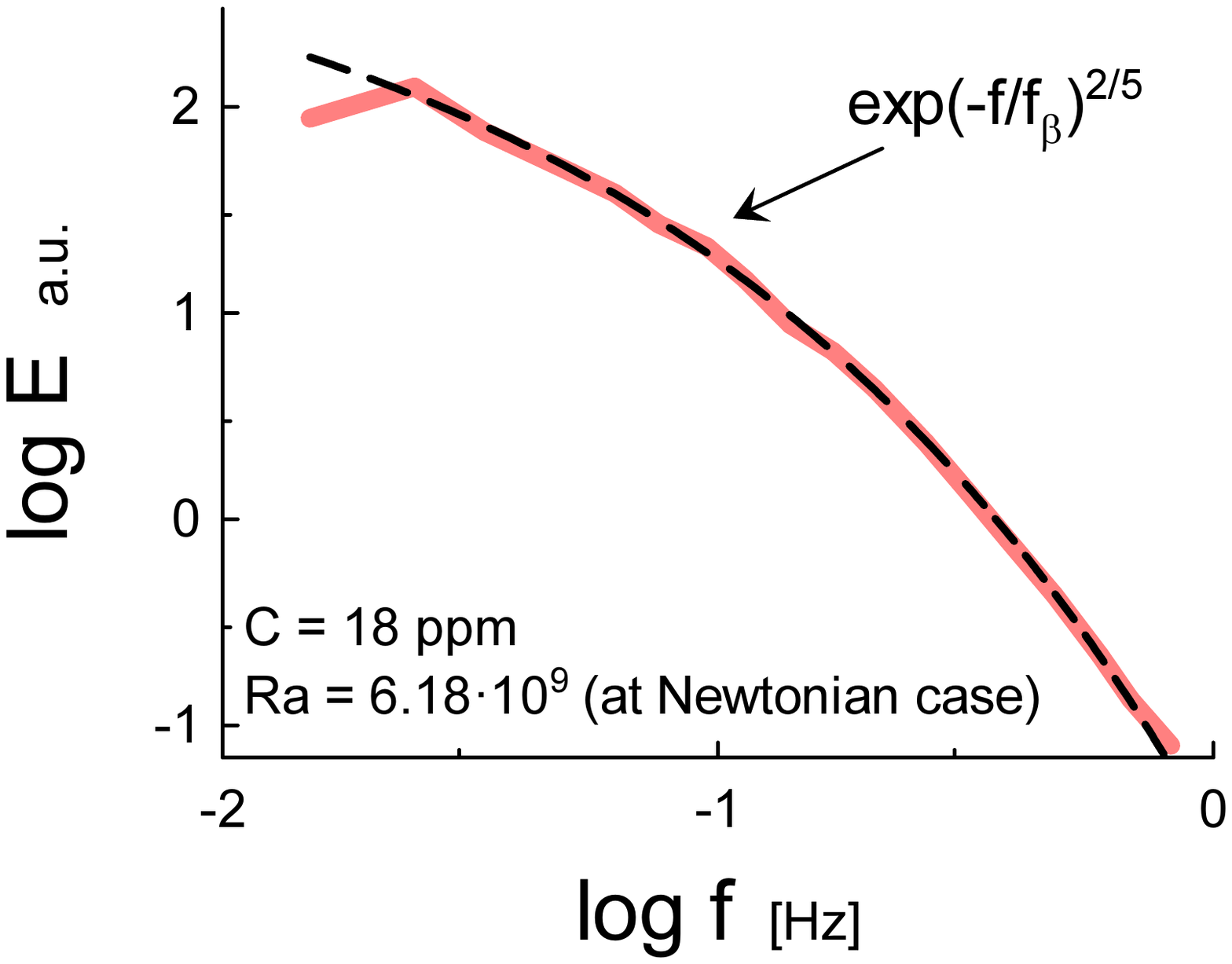}\vspace{-3.2cm}
\caption{\label{fig4} Vertical velocity power spectrum  for RBC with polymer additives. $Ra = 6.18 \cdot 10^9$ at the Newtonian case (without the polymer additives). The dashed line indicates the  spectral law Eq. (5) with $\beta = 2/5$. }
\end{center}
\end{figure}

  It is shown in the recent Ref. \cite{b3} that the spontaneous breaking of the relabeling symmetry dominates also the distributed chaos in fluid turbulence with polymer additives. Therefore, the main properties of this distributed chaos can be similar to the main properties of the RBC (without polymers) distributed chaos at the high Ra (including the enhancement of the heat transfer). Figure 4 shows vertical velocity power spectrum (data were taken from a RBC experiment reported in Ref. \cite{xx}) at a tiny polymer (Polyacrylimid in water) concentration $c=18$ ppm. Without the polymer additives (Newtonian case) Rayleigh number $Ra = 6.18 \cdot 10^9$. The measurements were produced in the center of the convection cell. The dashed line is drawn in order to indicate the spectrum Eq. (5) with $\beta =2/5$.

\section{Acknowledgement}

I thank X. He for sharing the data used in the Figs. 2 and 3.


\begin{thebibliography}{99}

\bibitem{bur}A. Burlot et al., J. Fluid Mech. {\bf 765}, 17 (2015).
\bibitem{b} A. Bershadskii, arXiv:1608.00489 (2016).
\bibitem{ll2} L.D. Landau and E.M. Lifshitz, Mechanics (Pergamon Press 1969).
\bibitem{saf} P. G. Saffman, J. Fluid. Mech. {\bf 27}, 551 (1967).
\bibitem{dav1} P. A. Davidson, J. Phys.: Conference Series {\bf 318} 072025 (2011).
\bibitem{dav2} P. A. Davidson P.A. Turbulence in rotating, stratified and electrically conducting fluids. (Cambridge University Press, 2013).
\bibitem{b1} A. Bershadskii, arXiv:1512.08837 (2015).
\bibitem{nrz} A. C. Newell, B. Rumpf, V. E. Zakharov, Phys. Rev, Lett., {\bf 108},
194502 (2012)
\bibitem{b2} A. Bershadskii, arXiv:1601.07364 (2016).
\bibitem{as} S. Ashkenazi and V. Steinberg, Phys. Rev. Lett. {\bf 83}, 3641 (1999).
\bibitem{my} A. S. Monin, A. M. Yaglom, Statistical Fluid Mechanics, Vol. II: Mechanics of Turbulence (Dover Pub. NY, 2007).  
\bibitem{mas} J. E. Marsden et. al., arXiv:math/0005034 (2000); J.Geom.Phys. {\bf 38}, 253 (2001).
\bibitem{y} A. Yahalom, arXiv:solv-int/9407001 (1994); J. Math. Phys. {\bf 36} 1324 (1995).
\bibitem{pm} N. Padhye and P. J. Morrison,  Phys. Lett. A {\bf 219}, 287 (1996).
\bibitem{fs} Y. Fukumotoa , H. Sakumab, Procedia IUTAM {\bf 7} 213 ( 2013) .
\bibitem{b3} A. Bershadskii, arXiv:1605.09291 (2016).
\bibitem{he} X. He, D. P. M. van Gils, E. Bodenschatz, and G. Ahlers, PRL {\bf 112}, 174501 (2014).
\bibitem{n}  J.J. Niemela, L. Skrbek, K.R. Sreenivasan, and R.J. Donnelly, J. Fluid Mech. {\bf 449},169 (2001).
\bibitem{he1} X. He, private communication. 
\bibitem{he2} X. He et. al, PRL {\bf 108}, 024502 (2012).
\bibitem{pz} I. Procaccia and R. Zeitak, Phys. Rev. Lett. {\bf 62}, 2128 (1989).
\bibitem{bnps} A. Bershadskii, J. J. Niemela, A. Praskovsky, and K. R. Sreenivasan, Phys. Rev. E {\bf 69}, 056314 (2004).
\bibitem{lt} E. Levich and A. Tsinober, Phys. Lett. A {\bf 93}, 293 (1983).
\bibitem{xx} Y-C. Xie et al., J. Fluid Mech.{\bf 784}, R3 (2015).





\end{thebibliography}
\end{document}